\documentclass[a4paper,11pt]{article}

\usepackage{amsmath, amscd, amssymb, amsthm}

\usepackage{array}

\usepackage{graphicx}
\usepackage{rotating}

\usepackage{listings}

\usepackage{bm}

\usepackage{fixme}

\usepackage{capt-of}
\usepackage{float}

\usepackage{color}

\definecolor{mauve}{rgb}{0.58, 0, 0.82}
\definecolor{dkgreen}{rgb}{0,0.6,0}

\begin{document}

\section*{The Imaging Computational Microscope}
\label{chap:computational_microscope}

E. Paxon Frady, William B. Kristan Jr.


\subsection*{Abstract}

The Imaging Computational Microscope (ICM) is a suite of computational tools for automated analysis of functional imaging data that runs under the cross-platform MATLAB environment (The Mathworks, Inc.).  ICM uses a semi-supervised framework, in which at every stage of analysis computers handle the routine work, which is then refined by user intervention. The main functionality of ICM is built upon automated extraction of component signals from imaging data, segmentation and clustering of component signals, and feature computation and visualization. Analysis of imaging data through ICM can also be done in real-time, enabling guided imaging experiments. ICM is built using MATLAB's object-oriented class design, which allows the tool to be used both as a graphic user interface (GUI) as well as in custom scripts. ICM is freely available under the GNU public license for non-commerical use and open source development, together with sample data, user tutorials and extensive documentation.

\subsection*{Introduction}
Large-scale recordings of neural data, such as calcium imaging or multi-unit electrode arrays, are necessary to understand the function of intact nervous systems. However, most experimentalists still rely on manual analysis techniques to synthesize and make sense of these ``big data'' experiments. As the number of neurons that can be simultaneously recorded continues to increase, manual analysis techniques will not be able to scale with the data and automated techniques will be necessary.

The Imaging Computational Microscope (ICM) is a computational tool for analyzing functional neural imaging data, such as calcium or voltage-sensitive dye imaging. The main functionality of ICM is built upon automated extraction of component signals from imaging data using Principal and Independent Component Analysis (PCA, ICA respectively; Mukamel et al. 2009; Hill et al. 2010), segmentation and clustering of component signals, and feature computation and visualization. These stages enable the experimentalist to start with raw imaging data and rapidly analyze large-scale data and produce high-level visualizations. The full set of computational analysis stages can be executed on the scale of minutes, which enables ICM to be used in real-time to guide experiments. 

\subsection*{Methods and Results}

\subsubsection*{Overview of Main Functions}
The Imaging Computational Microscope consists of several stages of analysis, each of which benefits from user interaction and inspection. Each stage is controlled using a series of tab menus (Fig. \ref{fig:labels_ss} R4), and ICM updates its displays based on the current stage and menu that is open. 

\textbf{Data.} In the data tab of the ICM, the user can browse the raw data, draw ROIs, perform motion correction, and enable concatenated-trial ICA (ctICA). In this stage, the Roi Editor (Fig. \ref{fig:labels_ss} R1) shows the raw imaging data, and ROIs can be manually drawn by clicking and dragging the mouse over the desired region of the image. The signals extracted by the ROI are plotted in the Data plot (Fig. \ref{fig:labels_ss} R6). These ROIs are stored as ``ROI sets'' and can be added or deleted using the ROI control list (Fig. \ref{fig:labels_ss} R3) as well as saved into and loaded from files. Multiple trials of data can be loaded into ICM and these can be browsed using the trial control list (Fig. \ref{fig:labels_ss} R2). 

\textbf{Pre-Processing.} In the pre-processing tab, the user can down-sample and smooth imaging data (see R4 in Fig. \ref{fig:pp_ss}). The ROI Editor will display the pre-processed data, and the extracted ROI signals of the pre-processed data will be plot in the Data plot (Fig. \ref{fig:labels_ss} R6).

\textbf{PCA.} In the PCA tab, the user can run Principal Components Analysis (PCA) on the imaging data, and browse the components. The principal components (PCs) are used to reduce the dimensionality of the imaging data and remove noise. When applied to imaging data, PCA produces a map (i.e. the spatial locations) and a source (i.e. the time-series) for each principal component. The PCs are typically combinations of many cellular signals or artifacts. Each PC map will be displayed as a frame in the ROI Editor, which can be browsed by clicking through the frames in the GUI. The corresponding source for each PC is plotted in the Component plot (Fig. \ref{fig:labels_ss} R7). 


\textbf{ICA.} In the ICA tab, the user can run and visualize Independent Components Analysis (ICA) of the imaging data (Bell \& Sejnwoski, 1995). ICA is run on a subset of the PCs, which is set by the user in the interface. The ICA algorithm also produces a map and a source for each of the independent components (ICs). The maps can be browsed as frames in the ROI Editor and the sources are plotted in the Component plot. The ICA algorithm has been shown to extract individual cellular signals, but it also extracts artifactual components. These must be viewed and sorted manually using the Remove box (see R4 in Fig. \ref{fig:ica_ss}). Further post-processing of the extracted components can also be performed using the controls in the ICA Post-Processing panel.


\textbf{Segmentation.} In the Segmentation tab, the user can segment the ICs and generate ROIs. The ROI Editor displays the results of the segmentation for each IC. The ROIs generated from segmentation are locked to their IC and can be used to quickly browse through all of the ICs.

\textbf{Visualization.} In the Visualization tab, ICM displays color-maps based on features computed from the IC sources. The user can manipulate the visualization settings using the controls in the Visualization Settings panel. The visualizations can be crafted using two built-in visualization algorithms, which are controlled in the Visualization panel (Fig. \ref{fig:labels_ss} R8). The algorithms assign each IC source 2-3 coefficients, based on some aspect of the component's activity. These coefficients are used to derive colors and create a colored activity map. Outside functions can also be used to create visualizations by assigning a color to each IC and setting these colors using the function \lstinline|set_viz_colors|.


We will illustrate the use and utility of ICM in more detail with two example walk-throughs of imaging data analysis.

\subsubsection*{PCA-ICA extraction of Calcium Imaging Data}
\begin{sidewaysfigure}[p]
  \centering
  \includegraphics[width=0.8\textwidth]{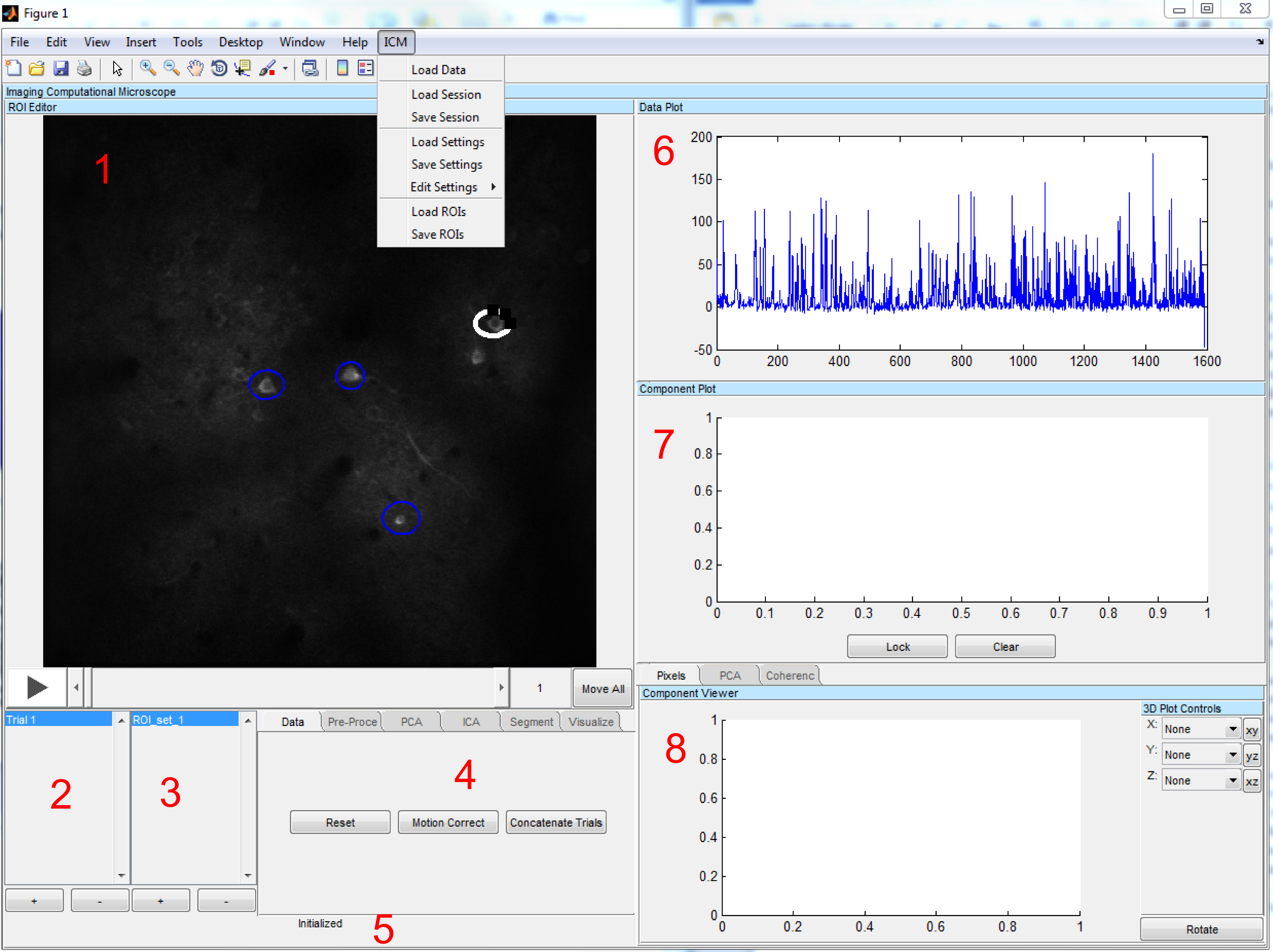}
  \caption{ICM screenshot with labeled regions.}
  \label{fig:labels_ss}
\end{sidewaysfigure}

ICM can load data from a .tiff or .mat file using the ICM menu in the interface (Fig. \ref{fig:labels_ss}), or data can be loaded into ICM through matlab functions \lstinline|set_data| or \lstinline|add_trial|. 
Data is loaded as a trial, and trials are managed using the trial management list (Fig. \ref{fig:labels_ss} R2).

The data can be browsed using the ROI Editor (Fig. \ref{fig:labels_ss} R1), which allows the user to draw manual ROIs and look through the frames of the imaging data. ROIs can be drawn by simply clicking and dragging the mouse on the image in the ROI editor. The signals from manually drawn ROIs are plotted in the Data plot (Fig \ref{fig:labels_ss} R6). Multiple sets of both manually and automatically generated ROIs can be managed using the ROI manager (Fig \ref{fig:labels_ss} R3).

Each stage of the analysis is set and controlled with a tab menu in the stage tab panel (Fig \ref{fig:labels_ss} R4). ICM displays different information depending on which stage of the analysis is being viewed. The current stage of the analysis is displayed beneath the stage tab panel (Fig \ref{fig:labels_ss} R5), which also indicates when ICM is busy computing for each stage.

\begin{sidewaysfigure}
  \centering
  \includegraphics[width=0.8\textwidth]{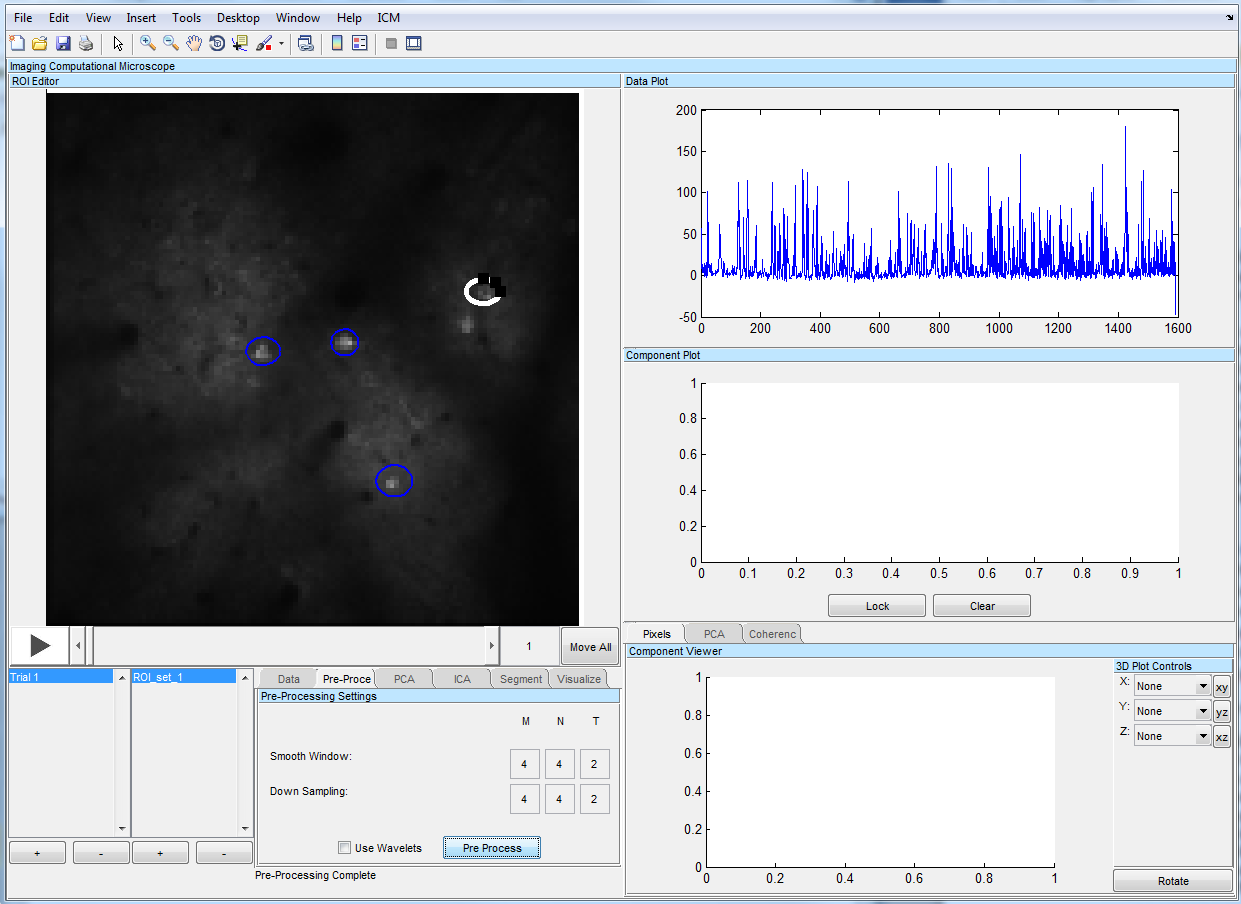}
  \caption{ICM screenshot in pre-processing stage.}
  \label{fig:pp_ss}
\end{sidewaysfigure}

In the next step, the preprocessing tab is opened, and the smooth window and down sampling values are set (Fig. \ref{fig:pp_ss}). The smooth window averages all pixels within a moving MxNxT window (rows, columns, time). The down sampling only keeps every MxNxT pixels, producing a smaller imaging data set. For this example, the original image data is 512x512x1600, but this is too large for the PCA-ICA analysis to be run on a desktop computer. We used a 4x4x2 smooth window and 4x4x2 down sampling to reduce the image size to 128x128x800. When the ``Pre Process'' button is pressed the original data is pre-processed and the pre-processed data is displayed in the ROI Editor. The original data can still be viewed in the data tab, and the pre-processing can be removed by clicking the ``Reset'' button in the Data tab.

\begin{sidewaysfigure}[p]
  \centering
  \includegraphics[width=0.8\textwidth]{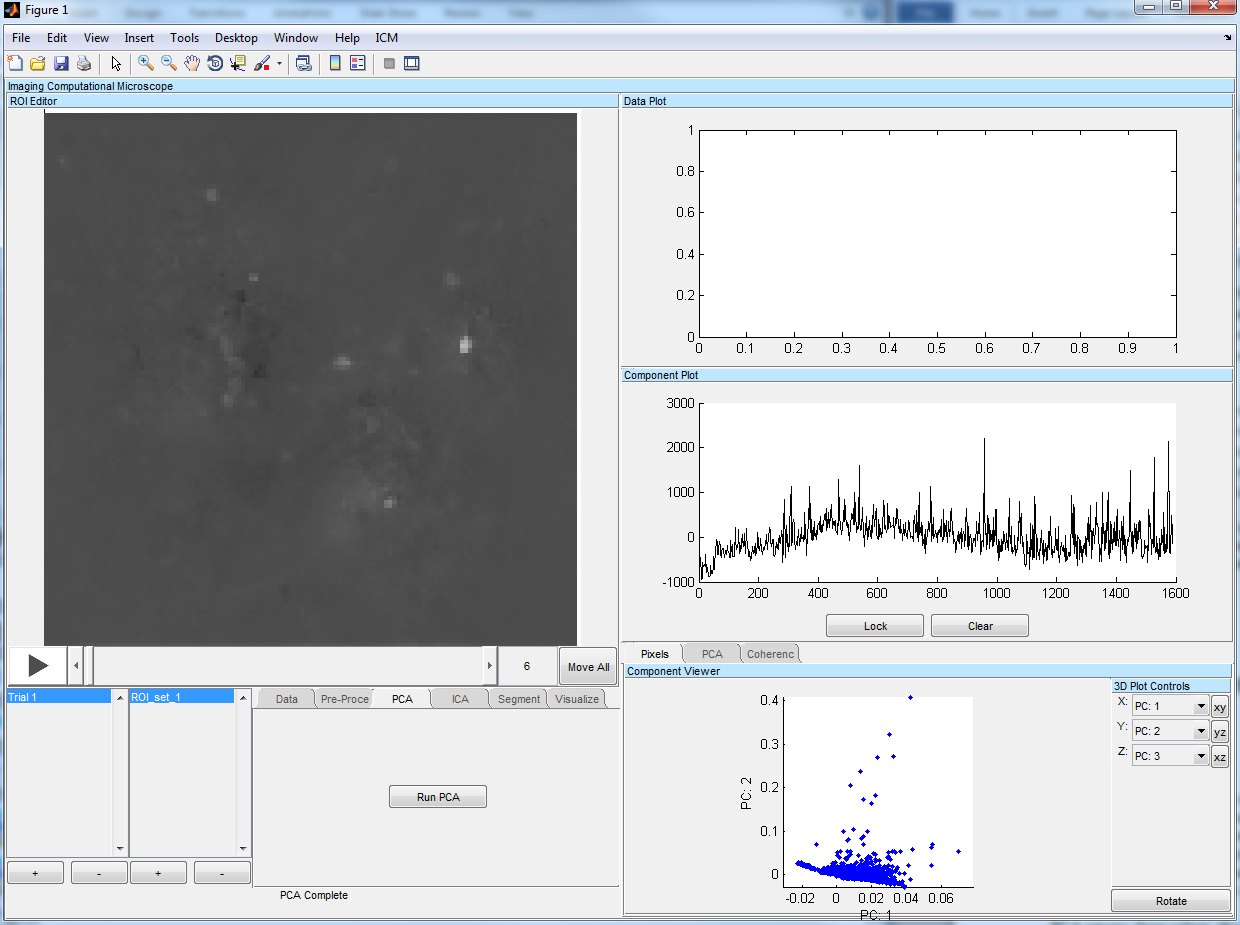}
  \caption{ICM screenshot in PCA stage.}
  \label{fig:pca_ss}
\end{sidewaysfigure}

Next, the principal components are computed. Every pixel in the preprocessed imaging data set is arranged in a matrix, where the rows are each pixels, and the columns are the values of the pixels over time. When the user clicks the ``Run PCA'' button (Fig. \ref{fig:pca_ss}), the principal components of this matrix are then computed, which decomposes the imaging data into several components, each of which has a ``source'' and a ``score''. Sources are the time series of the extracted components, and the scores indicate the coefficient of the source for each pixel. The scores are rearranged back into an image to produce a ``map'', which shows the spatial locations from which the sources are produced. The principal components are typically combinations of cellular signals, and do not reveal individual cells.

\begin{figure}[t]
  \centering
  \includegraphics[width=\textwidth]{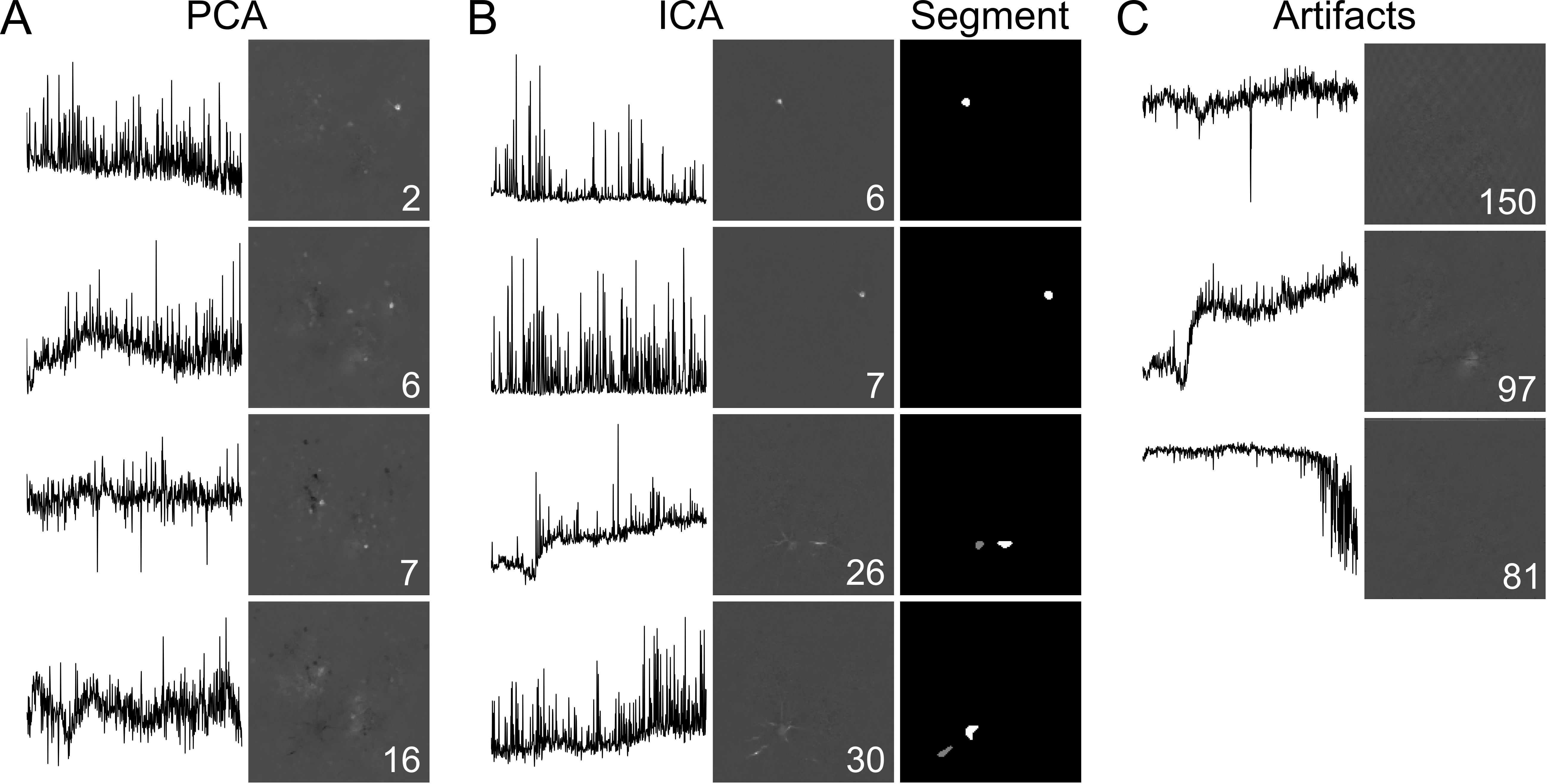}
  \caption[Example components]{\textbf{Example components.} (A) Four example principal components are shown. The PCs are typically mixtures of cells and artifacts. Sources (left) show calcium spikes, anti-spikes and background fluctuations, and maps (right) show a mixture of cells and background. (B) Four example independent component from cellular sources are shown. The bottom two components show the dendrites of two neurons that are out of the image plane (middle), but the Calcium spikes of these cells (left) can still be captured using the PCA-ICA extraction. The results of segmentation for each IC is shown on the right. (C) Three example independent components that are artifacts. The top shows corrigation due to image sampling artifact, the middle is background fluorescence, and the bottom is an edge artifact.}
  \label{fig:walk1_examples}
\end{figure}

The principal components can be browsed in the ROI editor with the PCA tab open. Each frame in the ROI Editor shows the map of a different principal component, and the source of each component is plotted in the Component Plot. Several more example principal components can be seen in Figure \ref{fig:walk1_examples}A. 

\begin{sidewaysfigure}[p]
  \centering
  \includegraphics[width=0.8\textwidth]{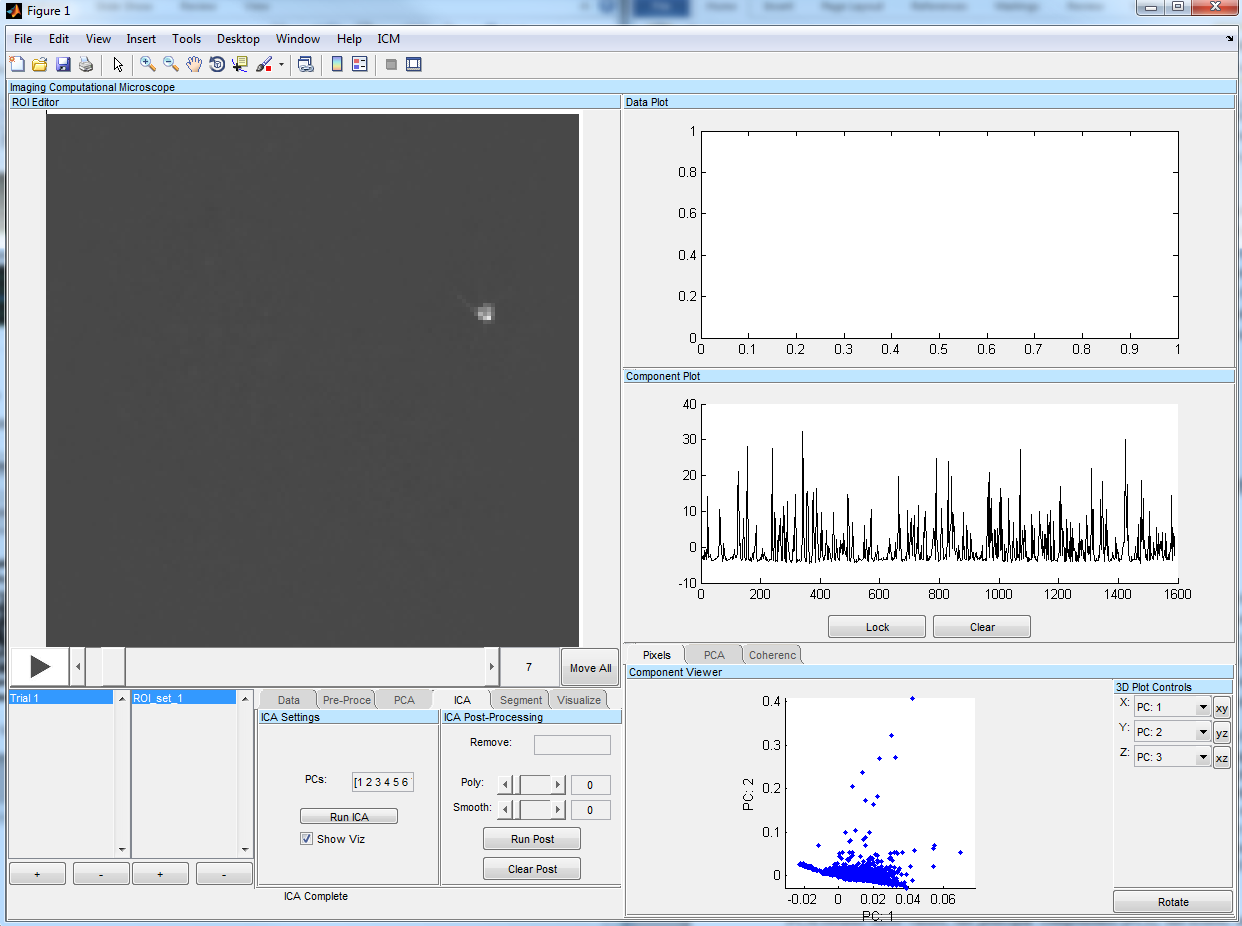}
  \caption{ICM screenshot in ICA stage.}
  \label{fig:ica_ss}
\end{sidewaysfigure}

To reveal individual cells, the independent components are then computed from a subset of the principal components. Typically the top $N$ principal components are used, where $N$ is slightly larger than the number of neurons being recorded from. This can be set using the ``PCs'' edit box in the ICA tab (Fig. \ref{fig:ica_ss}). The ICA algorithm will attempt to find the same number of independent components as principal components included, and so there should be at least as many PCs used as cells. Typically, more PCs are needed than cells, because many independent components are extracted that correspond to motion, background, bleaching, or other artifacts.

Once the PCs are chosen and ``Run ICA'' is clicked, ICM computes the independent components using the fastica algorithm by default. Further changes to the fastica settings can be made through a settings struct that can be changed programatically (see Documentation), as well as choosing different ICA algorithms (e.g. infomax (Bell \& Sejnowski, 1995) and spatio-temporal ICA (Mukamel et al. 2009)). Like  PCA, ICA also produces sources and scores, where the sources correspond to the independent signals and the scores describe which pixels are contributing to the source. The scores are rearranged back into an image to produce a map, and the maps show the spatial locations of the independent components. 

Several example independent components are shown in Figure \ref{fig:walk1_examples}B and C. The ICA algorithm pulls out both components that are cellular signals (Fig. \ref{fig:walk1_examples}B) as well as components that are artifacts (Fig. \ref{fig:walk1_examples}C). These must be sorted manually using the interface, and components which are artifactual can be removed from further analysis by adding them to the ``Remove'' edit box. Further post-processing can be performed on the ICs using the tools in the ``ICA Post-Processing'' panel.

\begin{sidewaysfigure}[p]
  \centering
  \includegraphics[width=0.8\textwidth]{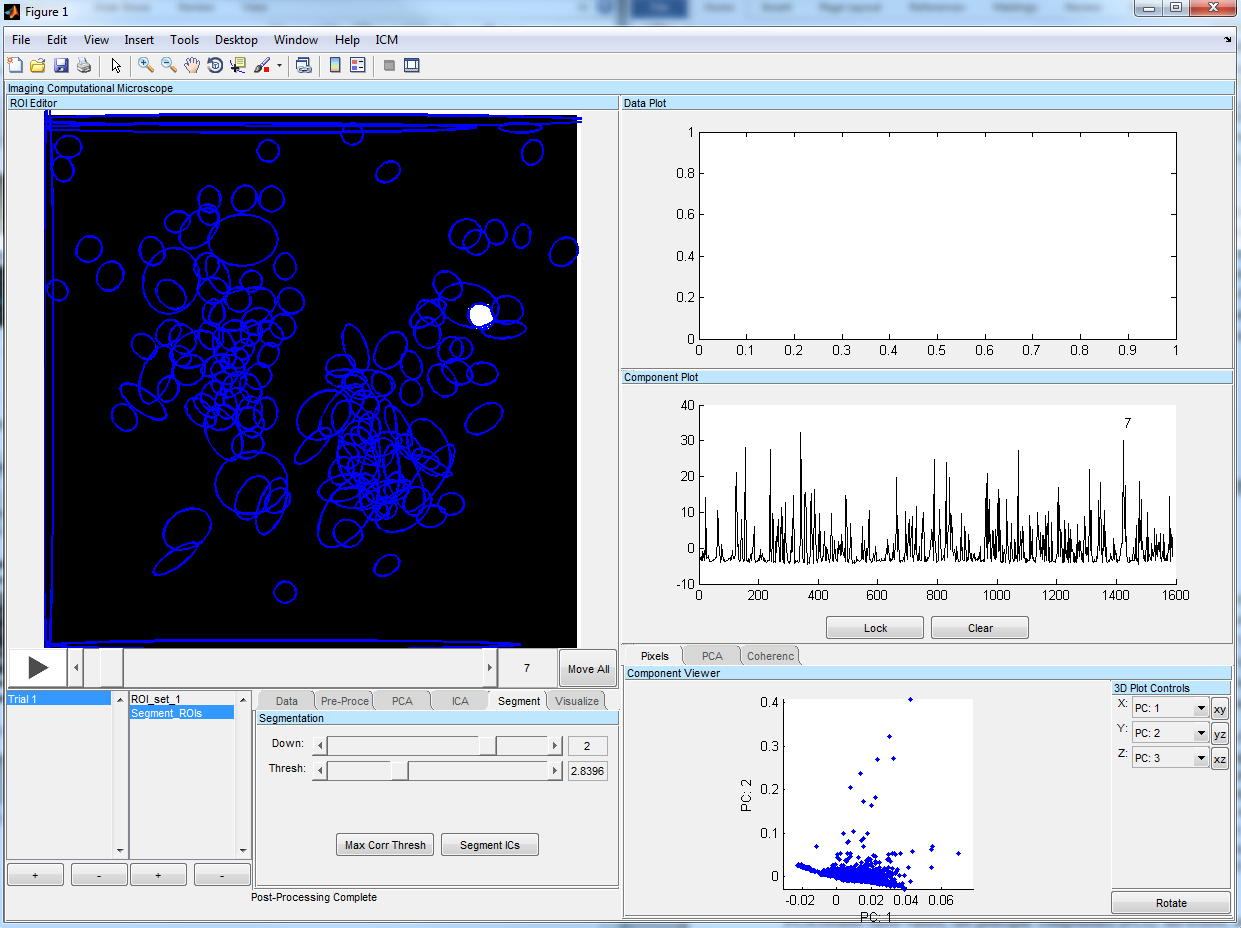}
  \caption{ICM screenshot in segment stage.}
  \label{fig:segment_ss}
\end{sidewaysfigure}

Regions-of-interest can then be automatically generated from the ICA maps. To compute the regions of interest, ICM uses a threshold to segment the IC maps. In the Segment tab (Fig. \ref{fig:segment_ss}), the threshold level and amount of down-sampling are set for the segmentation algorithm. When the ``Segment ICs'' button is pressed, binary masks are created for each IC, which are displayed in the ROI Editor when the Segment tab is open. Each contiguous region of the binary mask is then matched with the best fitting oval to produce the ROI. The ROIs produced are saved in the ROI manager (Fig. \ref{fig:labels_ss} R3). 

ROIs produced automatically remain linked with their corresponding IC and this allows the user to browse the ICs by clicking on the corresponding ROI. The ROIs also show the user the spatial localization and extent of all of the ICs. 

Each IC can potentially have multiple ROIs, because there can be multiple spatially isolated regions for a single IC. In this example data, several cell soma's are slightly misaligned with the image plane, but their dendritic branches are still recorded (see IC 26 and 30 at bottom of Fig. \ref{fig:walk1_examples}B). Because of this, the segmentation algorithm breaks up these cells into multiple ROIs (Fig. \ref{fig:walk1_examples}B, right). 

A major advantage of the PCA-ICA extraction is that this algorithm does not depend on spatial-localization to extract component signals. Calcium signals from cells or axons that do not have a single localized spatial region would be virtually impossible to extract if only ROIs were used. The ICA component decomposition does not depend on spatial localization, which allows for clear signals to be extracted from out of focus cells or long axons (see bottom two examples in Fig. \ref{fig:walk1_examples}B). This suggests that many cells may be missed entirely when relying on ROI methods and that the ICA algorithm can get much higher signal-to-noise ratio under certain imaging conditions. Further, ICA can separate components that have overlapping spatial locations. For example, IC 150 (Fig. \ref{fig:walk1_examples}C) shows clear corrugations due to artifacts of the imaging acquisition system. These striations cover the entire image, but ICA can separate this artifact from the cellular components because the pixels share statistical patterns caused by the artifact. It would be impossible for ROI methods to separate spatially overlapping components.

\begin{sidewaysfigure}[p]
  \centering
  \includegraphics[width=0.8\textwidth]{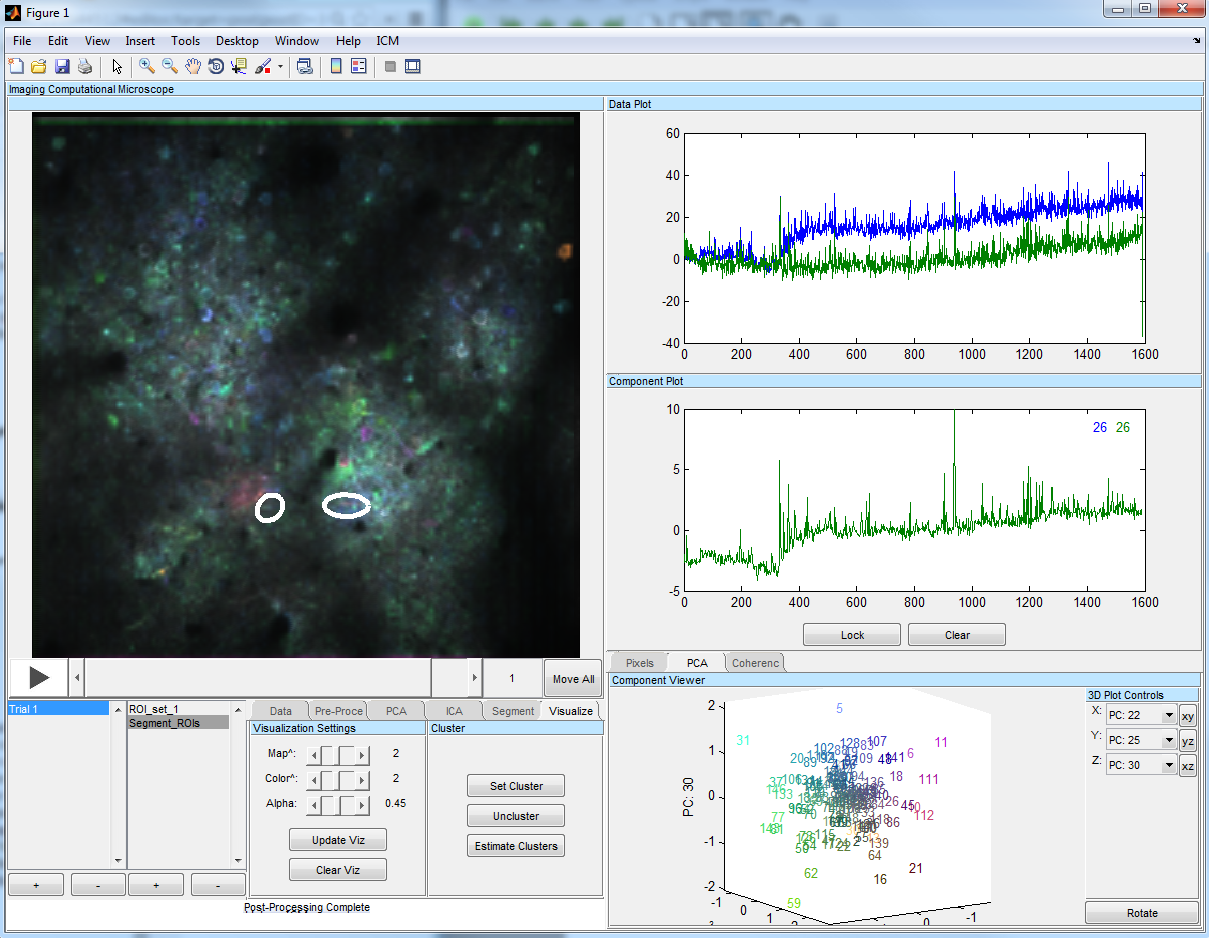}
  \caption{ICM screenshot in visualization stage.}
  \label{fig:viz_ss}
\end{sidewaysfigure}

Finally, visualizations are created in the visualization stage. Two simple visualization systems are built into ICM, and here we will illustrate the principal component visualization of the IC sources. The PCA component viewer can be opened in the visualization control tabs (Fig \ref{fig:labels_ss}. R8), and this allows the user to select three principal components to visualize. Each independent component has coefficients in the principal component space, and the user selects which three dimensions to view. The component maps are then colored based on the coefficients of the ICs given the dimensions chosen, and these are overlayed on the image data to create a visualization of activity. The settings of the visualization can be manipulated using the controls in the ``Visualization Settings'' panel (Fig. \ref{fig:viz_ss}).

\begin{figure}[t]
  \centering
  \includegraphics[width=\textwidth]{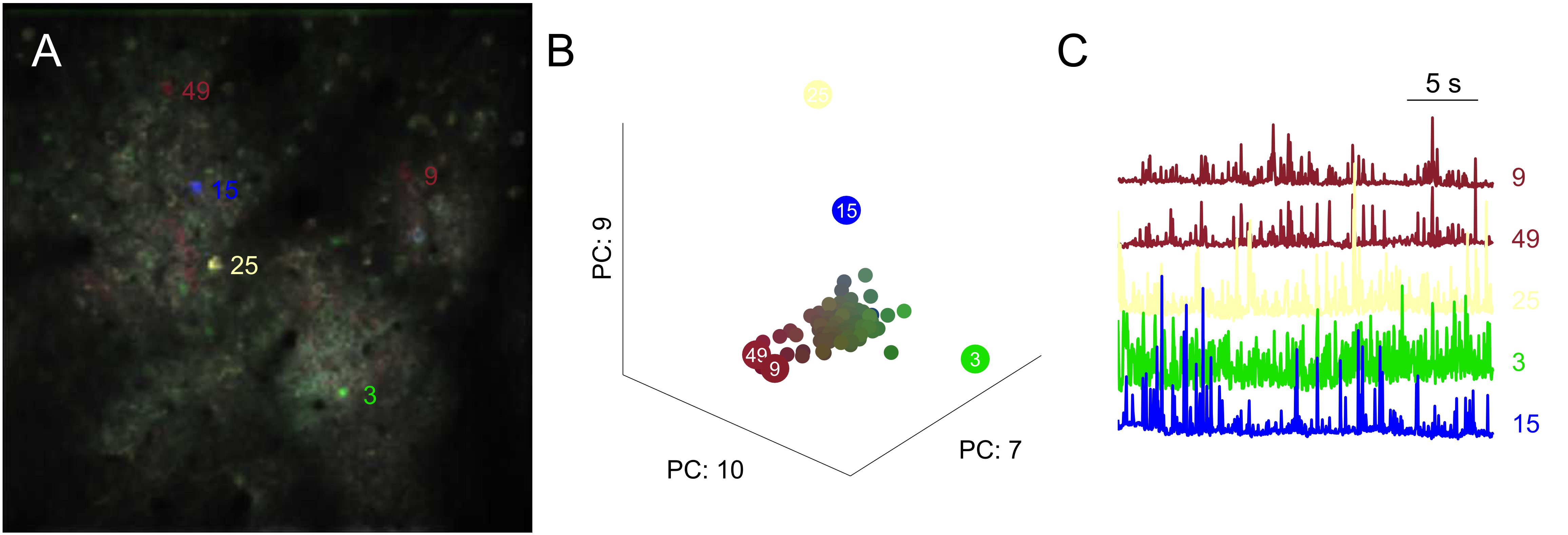}
  \caption[ICM visualization reveals patterns in large-scale imaging data.]{\textbf{ICM visualization reveals patterns in large-scale imaging data.} (A) A visualization of cellular activity is created by ICM by coloring IC maps based on the PC coefficients. Five cells are indicated with numbers as examples. Cells 9 and 49 are nearly the same color. (B) Three principal components are chosen in the GUI and the coordinates of each IC are plotted in three dimensions. The colors for each IC are based on these coordinates. Five examples are labeled. (C) The sources of the five example components are shown. Components 9 and 49 have very similar spiking activity, which is reflected by their similar color. The other cells have different spiking activity and are colored differently.}
  \label{fig:pc_ica_viz}
\end{figure}

Visualizations are extremely useful for quickly assessing activity patterns in imaging data. In Figure \ref{fig:pc_ica_viz}A, we illustrate the PCA visualization and highlight some example components of interest. To create this visualization, PC 7, 10, and 9 were selected in the PC component viewer (Fig. \ref{fig:labels_ss} R8) and ICM uses the coefficients from these three coordinates to create the visualization (Fig. \ref{fig:pc_ica_viz}B). This particular example shows a handful of cells with different colors, and notably IC 9 and 49 are very close to the same color. This is the result of the fact that cells 9 and 49 have very correlated spiking activity (Fig. \ref{fig:pc_ica_viz}C), which places them in nearly the same location in PCA space (Fig. \ref{fig:pc_ica_viz}B).  

\subsubsection*{Multi-trial extraction of Voltage-Sensitive Dye Imaging Data}

Often many different trials are performed while imaging under different experimental conditions, and it is important to extract the same cells under these different conditions for analysis. The best solution for this situation is concatenated-trial ICA, where the multiple imaging trials are aligned and concatened into a single imaging data set. This extracts the same cell as the same component from each trial, ensuring that cross-trial comparisons are valid. 

In this section, we illustrate the programmatic use of ICM with an example of concatenated trial ICA (ctICA). This walk-through follows the script walkthrough2.m in Supplementary Appendix B. This will illustrate scripted use of ICM, and reveal some of the more advanced features of the software. We will follow along the Matlab script and also point out how actions in the script can be done in the GUI.


To enable ctICA, several data files would first be loaded through the GUI menu or using the function \lstinline|add_trials| (Supplementary Appendix B Section 2). These trials would be imaging data collected from the same region of the brain and the changes in position of the image field or its z-depth must be minimal.

To perform ctICA, ICM concatenates multiple trials as if they were a single imaging acquisition. For this to work, however, the pixels must be consistently aligned with the sources of the signals. ICM uses an image registration algorithm (Evangelidis \& Psarakis, 2008) to align the acquisitions across trials. To enable ctICA, the ``Concatenate Trials'' button is pressed in the Data tab when ICM is in the ``Initialized stage''. This calls the function \lstinline|align_trials|, which performs the image registration on all of the trials currently loaded and puts ICM into ctICA mode (Supplementary Appendix B Section 3).

Once ICM is in ctICA mode, the process of extracting the independent components (ICs) is essentially the same as the single trial extraction. Pre-processing, PCA, and ICA are run using the controls in each of their tabs. These VSD recordings are much less sensitive than Calcium imaging recordings, and to maximize the signal we set the pre-processing smooth window to 6x6x1, which is about the size of the smallest cell. This smoothing makes each pixel record the average of the 6x6 pixels around it, increasing the signal-to-noise ratio of the voltage signals, which will improve the extraction of the ICs (Supplementary Appendix B Section 4).

\begin{figure}[t]
  \centering
  \includegraphics[width=0.7\textwidth]{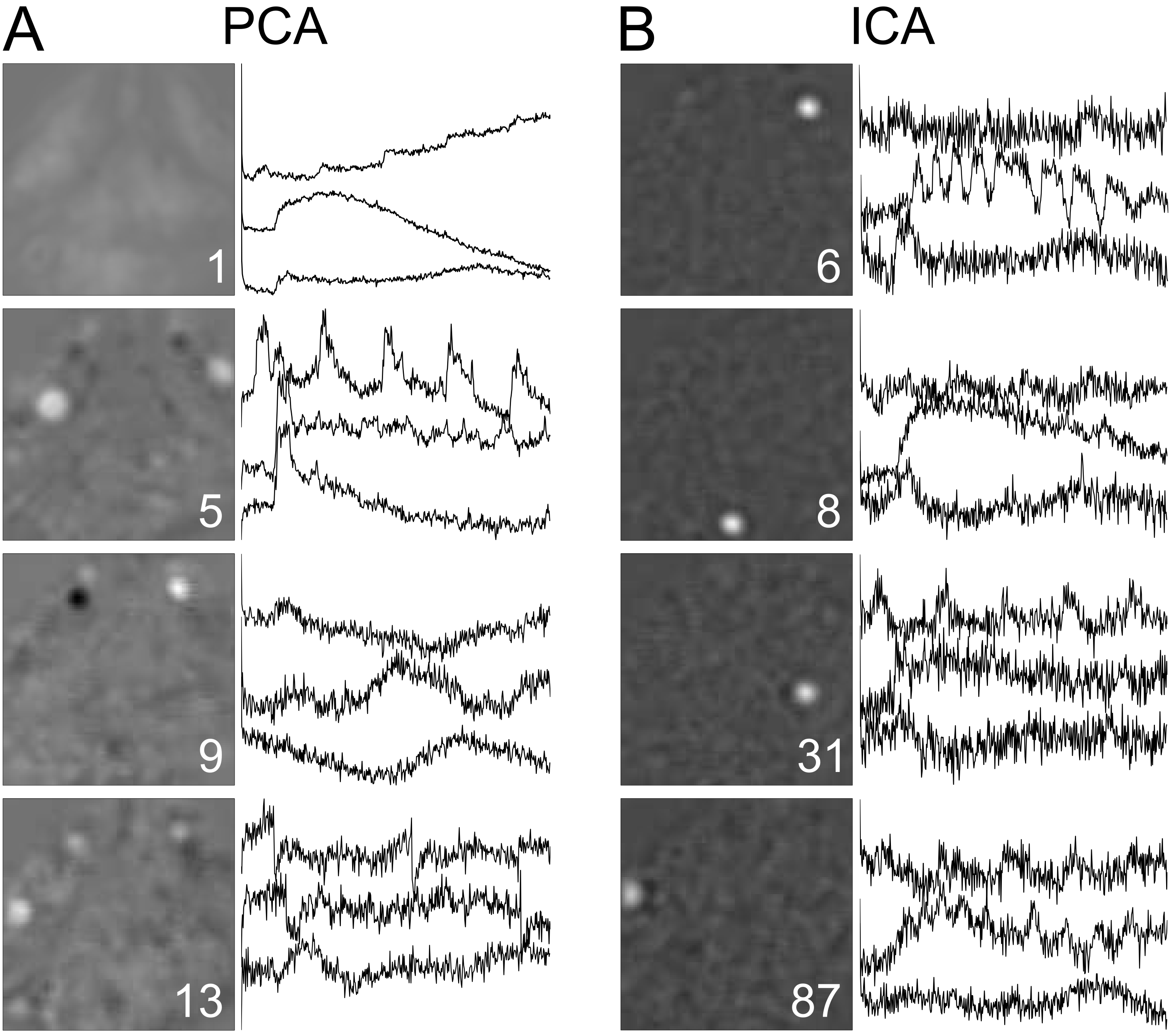}
  \caption[Examples of ctICA components.]{\textbf{Examples of ctICA components.} (A) Four example principal components from ctICA analysis. Three sources (right) are shown corresponding to signals from the three concatenated trials. The component number is indicated at the bottom right of the component maps (left). (B) Four example independent components. The three sources (right) shows activity from an individual cell across three different trials. The component number is indicated at the bottom right of the maps (left).}
  \label{fig:walk2_examples}
\end{figure}

PCA is then run on the concatenated data as if they were a single acquisition. Each PC will have a map and a source, but the source is now a time-series that extends through all three concatenated trials. The map is the same for all three trials. After PCA is run, the maps and sources of each trial can be viewed in ICM. The single concatenated source is split back up to individual trial sources, and these can each be seen by looking at the PCA tab in the different trials using the trial selection list (Fig. \ref{fig:labels_ss} R2). In Figure \ref{fig:walk2_examples}A, four example PCs are shown along with the three sources from the individual trials (Supplementary Appendix B Section 5).

ICA is then run using the PCs defined in the settings struct. The number of PCs to use in ctICA should be even larger than in individual trials, because more artifactual components may be present due to distortions when concatenating the trials. Again, the ICA maps are the same across all three trials, but the sources will be unique for each trial. This means that the locations from which the sources come from are the same (i.e. the component is the same cell across trials), but the cell signals are different (because the cell's activity is different in different trials). These maps and sources for each trial can be viewed using the trial selection list (Supplementary Appendix B Section 6).

\begin{figure}
  \centering
  \includegraphics[width=0.5\textwidth]{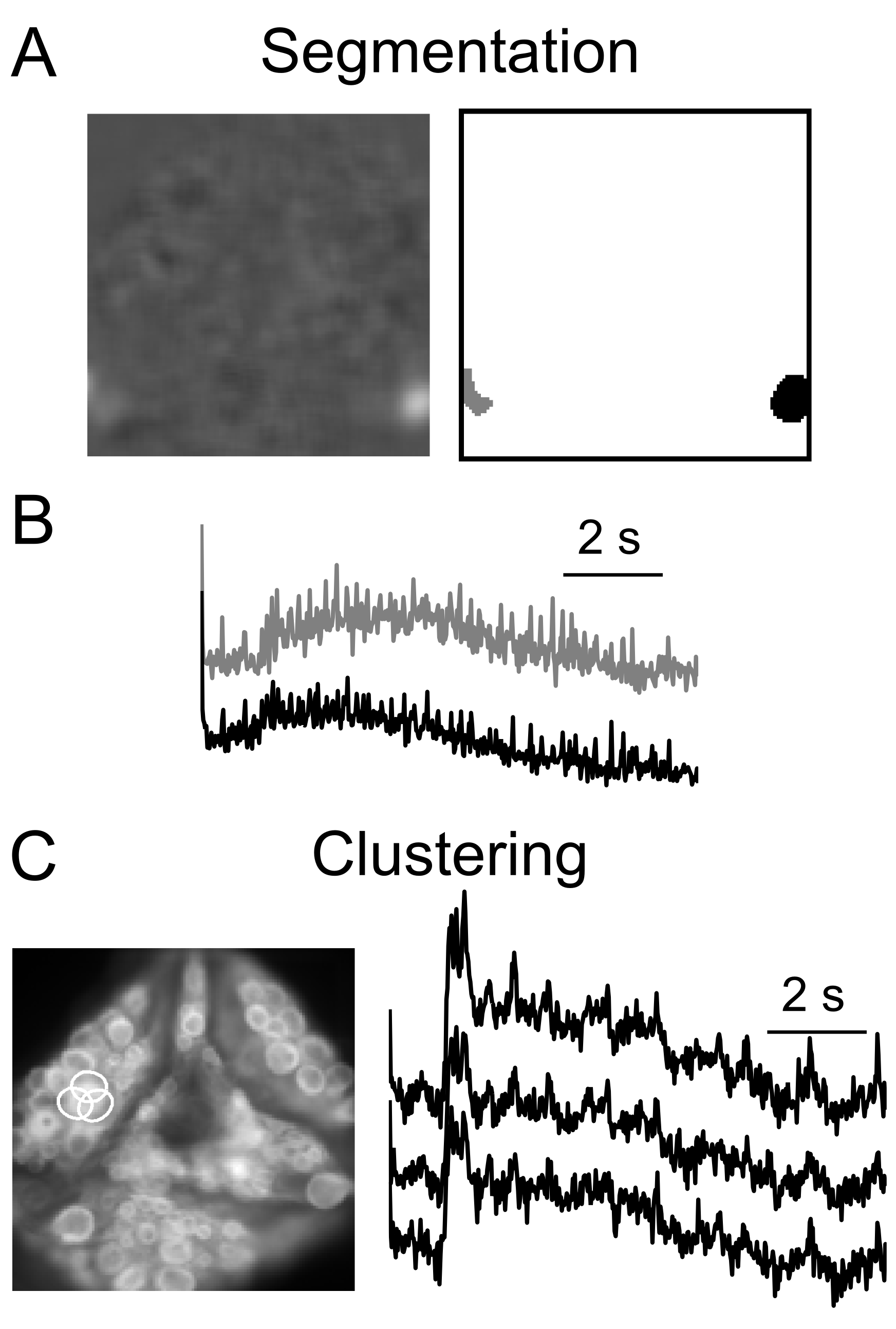}
  \caption[Segmentation and Clustering]{\textbf{Segmentation and Clustering of ICs} (A) An example IC where two correlated cells are extracted as the same component (left) can be spatially segmented with a threshold (right). (B) The ROI signals from the gray and black segments derived in panel A show these two cell's spiking activity is highly correlated. (C) An example of three extracted ICs that correspond to a single cell. These can be manually clustered using the GUI.}
  \label{fig:walk2_segment_cluster}
\end{figure}

Segmentation is run in the same fashion as the individual trials. Since the segmentation is done on the ICA maps and these are the same across concatenated trials, there will be the same segments for all trials (Supplementary Appendix B Section 7). Segmentation is needed when there are cellular signals that are highly correlated, because the ICA algorithm can lump strongly correlated signals into a single component. In Figure \ref{fig:walk2_segment_cluster}A an IC map is shown for a component that captures the strongly correlated activity of two electrically coupled bilateral pairs. These cells can easily be segmented because they are spatially separated (Fig. \ref{fig:walk2_segment_cluster}A, right). 

The ICA algorithm can also split a single cell source into multiple components (Fig. \ref{fig:walk2_segment_cluster}C). This sometimes occurs with very large cells. The user can manually set components into a cluster by selecting the ROIs of the ICs and pressing ``Set Cluster'' in the Cluster control panel in the Visualize tab (see Fig. \ref{fig:viz_ss}). 

Finally, visualizations can be made for each trial using the concatenated results. In the first trial, a cell was stimulated with electrodes at 0.5 Hz while VSD activity was simultaneously recorded. ICM's coherence tools are useful for creating a visualization of rhythmic signals captured by the VSD imaging. To create this visualization, first the IC of the stimulated cell was selected by browsing the ICs using the GUI (Fig. \ref{fig:ctICA_viz}A top trace), which shows the spiking of the stimulated cell. This was set as the ``base'' in the coherence tool, which then calculates the coherence between the base and the other VSD signals (Bokil et al. 2010; chronux.org). The frequency of the coherence was then set to the stimulus frequency, and the coherence magnitude and phase at that frequency for every IC is plotted as a polar plot (Fig. \ref{fig:ctICA_viz}B). The ICs with a coherence magnitude greater than the significance level (dashed red line in Fig. \ref{fig:ctICA_viz}B) are given a color based on their phase. This color is used to color the IC maps and create a visualization of the activity (Fig. \ref{fig:ctICA_viz}C) (Supplementary Appendix B Section 8).

Note that because the fastica algorithm uses a random initialization, it will not always return the exact same components or order of components. In the script, we include some advanced commands to initialize the mixing matrix, which should then lead to the exact same components being computed. If the script is not producing the correct visualizations, then it is possible that the component order of the ICs is incorrect and the settings of the visualizations are also incorrect. It is typical to have to browse the ICs after they are computed and search for the cellular signals of interest by hand. 

\begin{figure}
  \centering
  \includegraphics[width=\textwidth]{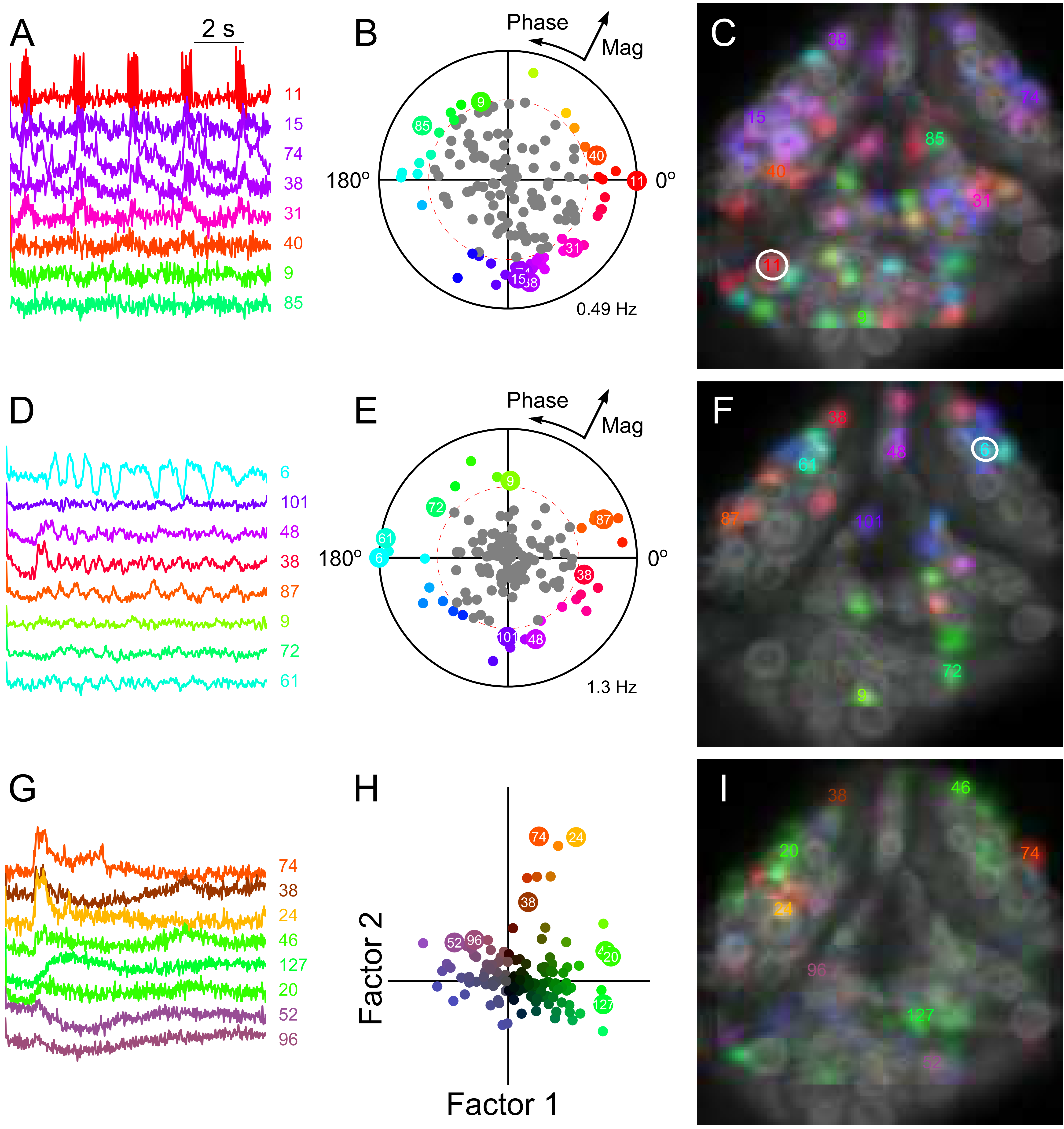}
\end{figure}

\begin{figure}
  \captionof{figure}[Visualization of multiple trials with ctICA.]{\textbf{Visualizations of multiple trials with ctICA.} (A) A sensory P cell was stimulated (top trace) and several VSD traces of example cells are shown which respond to the P cell stimulus. (B) The coherence of the VSD traces was computed against the P cell trace. Cells which had a significant coherence (red dashed line) were assigned a color based on the phase of their oscillation. (C) The IC maps are colored and an activity map is created to visualize neural responses. The white circle indicates the stimulated cell. (D-F) Same as A-C, except the coherence magnitude and phase of an oscillatory motor pattern, swimming, are calculated. (G) Several VSD traces of the shortening behavior. (H) The shortening responses are decomposed into two factors and the coefficients for each IC are plotted. These coefficients determine the color assigned to each IC. (I) An activity map is generated using the colors derived in panel H. }
  \label{fig:ctICA_viz}
\end{figure}

In the second trial, the leech is performing the swim behavior, which is a cyclical oscillatory behavior and many neurons can be seen oscillating in the VSD recordings. This can again be visualized using ICM's coherence tools. A known cell that shows strong oscillations is chosen manually as the base (Fig. \ref{fig:ctICA_viz}D top trace), and the dominant frequency of the oscillation is chosen as the coherence frequency (Fig. \ref{fig:ctICA_viz}E). In this case, the base cell oscillation is defined as the $180^\circ$ phase, and so the phases of all the other components are rotated. This is used to create a visualization of the oscillations of the neurons (Fig. \ref{fig:ctICA_viz}F) (Supplementary Appendix B Section 9).

In the third trial, we use an outside function to create the visualization colors and import these colors into ICM. This trial is the shortening behavior and we use factor analysis to find low-dimensional descriptors of the IC sources (Fig. \ref{fig:ctICA_viz}G). Factor analysis fits pre-defined curves to each IC source and returns the coefficients of these fits (Fig. \ref{fig:ctICA_viz}H). A color for each IC is then computed based on these coefficients and these are imported into ICM using the function \lstinline|set_viz_colors|. ICM then creates the visualization with the given colors (Fig. \ref{fig:ctICA_viz}I; Supplementary Appendix B Section 10).

\subsection*{Performance of ICA vs. ROI}
Regions-of-interest (ROIs) are classically used to extract signals from imaging data. An intrinsic issue with this method is that often the signals of interest will overlap and a single pixel can be receiving signal from multiple sources. This type of scenario is impossible to compensate with ROI methods, and some signal must be sacrificed or lost. An important advantage of the ICA algorithm is that it can use the higher-order statistics of the pixels to extract component signals and does not depend on spatial localization. This means that components that overlap on the same pixels can be separated by ICA, and that the ICA algorithm could potentially produce more accurate signals than ROIs. 

The ICA algorithm has been previously used to extract optical signals, and reliable cell signals can be extracted (Mukamel et al. 2009; Hill et al. 2010). However, the objective quality of this procedure has never been measured nor compared to other techniques for extracting optical signals. This is because for most imaging data, the imaged signals are based on Calcium, and there is no underlying ground-truth measure of Calcium. However, for VSD imaging, we do have an objective ground-truth measure: the intracellular potential, and can compare optical signals to intracellular recordings by performing simultaneous VSD imaging and electrophysiological recordings.

\begin{figure}
  \centering
	\includegraphics[width=0.5\textwidth]{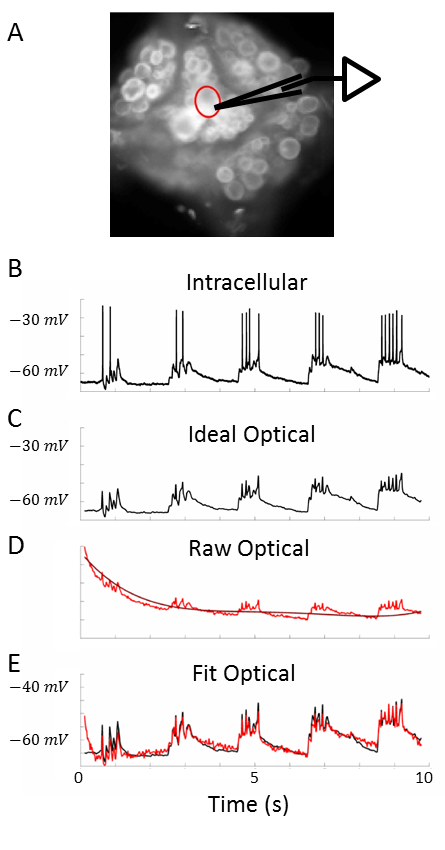}
  \caption[Calculation of Signal-to-Noise Ratio (SNR).]{\textbf{Calculation of Signal-to-Noise Ratio (SNR).} (A) Simultaneous optical and electrophysiological recordings of several cells under several different conditions 
(B) Raw intracellular voltage trace. 
(C) The Voltage trace is down-sampled to the optical frequency based on a method that mimics the optical sampling mechanism. 
(D) The raw optical trace is shown in red and a polynomial is used to fit the optical trace to the ephys trace (dark red). 
(E) The fit optical trace is overlayed on the ideal optical trace. 
(F) The SNR is calculated as the standard deviation of the Ideal trace divided by the standard deviation of the difference between Ideal and Fit (i.e. the residuals).}  
  \label{fig:snr_calculation}
\end{figure}

To assess the relative advantages of ICA based signal extraction, we compared the quality of signals extracted by ICA to those derived by the classical region-of-interest (ROI) technique. We made simultaneous intracellular and optical VSD recordings (Miller et al. 2012) of several cells in the leech ganglion under several different conditions (Fig. \ref{fig:snr_calculation}A). We came up with an objective measure of performance by computing the ``signal-to-noise ratio'' (SNR) of the extracted optical signal with the electrophysiological signal. The most desired result of VSD imaging would be that the optical signals perfectly mimic the intracellular recordings.  However, the sampling frequency is different between the intracellular and optical recordings, and optical signals are corrupted by bleaching artifact. To correct for these expected downsides of imaging, we down-sampled the intracellular recording (Fig. \ref{fig:snr_calculation}B) to the optical frequency. The down-sampling averages all of the intracellular recording samples that occur during the exposure for each frame of the imaging data, mimicking the way the optical signal is sampled. This results in the Ideal Optical signal (Fig. \ref{fig:snr_calculation}C), which would be the best optical signal we could possibly record given our imaging sampling methods. 

\begin{figure}[t]
  \centering
  \includegraphics[width=0.7\textwidth]{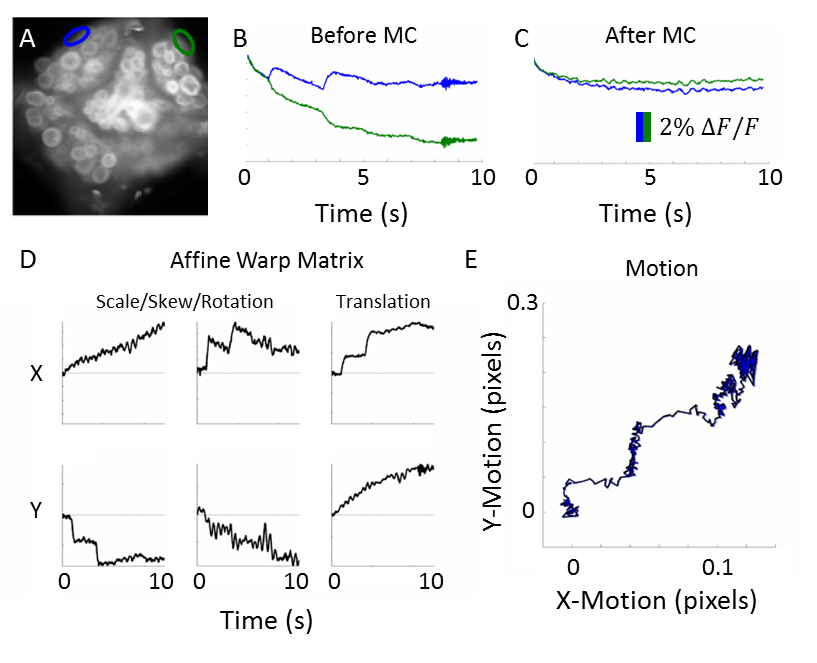}
  
  \caption[Sub-pixel motion correction with ECC.]{\textbf{Sub-pixel motion correction with ECC.} (A) Raw image of ganglion stained with VSD with two ROIs at the edge of the ganglion. (B) Sub-pixel motion artifact is apparent and large compared with VSD signals. (C) Motion artifact can be removed with ECC image registration algorithm. (D) The registration algorithm performs an affine transformation of the pixels. Each panel corresponds to the warp matrix values over time. (E) The resulting motion artifact that is removed. The separation of blue and black highlights the impact of the skew and rotation values of the affine transformation.}
  \label{fig:motion_correction}
\end{figure}

To remove the bleaching artifact, we use a polynomial to fit the Raw Optical trace to the Ideal Optical signal (Fig. \ref{fig:snr_calculation}D). Figure \ref{fig:snr_calculation}E shows the Fit Optical signal overlayed onto the Ideal Optical signal. The SNR is computed as the standard deviation of the Ideal Optical signal divided by the standard deviation of the difference between the Ideal and the Fit (Fig. \ref{fig:snr_calculation}F). 

\begin{equation}
SNR = \frac{\sigma(Ideal)}{\sigma(Ideal - Fit)}
\end{equation}

Motion artifact is another big component of errors in optical signals. ICM includes a built-in motion-correction algorithm (Evangelidis \& Psarakis, 2008). This simply registers each frame of the movie to the previous frame using an affine transform (i.e. scale, skew, rotation and translation). This algorithm can even remove sub-pixel motion artifact, which cannot be seen by eye. Figure \ref{fig:motion_correction} illustrates sub-pixel motion correction removed by the algorithm. The blue and green ROIs shown in Figure \ref{fig:motion_correction} are at the edge of the ganglion, and they reveal the subtle motion artifact (Fig. \ref{fig:motion_correction}B) as two slight tugs towards the blue ROI. This motion can be fairly well removed using the motion correction algorithm (Fig. \ref{fig:motion_correction}C), which computes the Affine Warp matrix for each frame to remove the motion (Fig \ref{fig:motion_correction}D, E).

We assessed the performance of ICA and ROI extraction techniques by comparing the SNR of several simultaneous recordings. We also compared the two techniques with and without sub-pixel motion correction (no recordings with significant motion were used). Fifteen recordings were made during several different conditions: spontaneous activity, chemical pre-synaptic stimulation, electrical pre-synaptic stimulation, and the swimming behavior. 

\begin{figure}[t]
  \centering
  \includegraphics[width=0.8\textwidth]{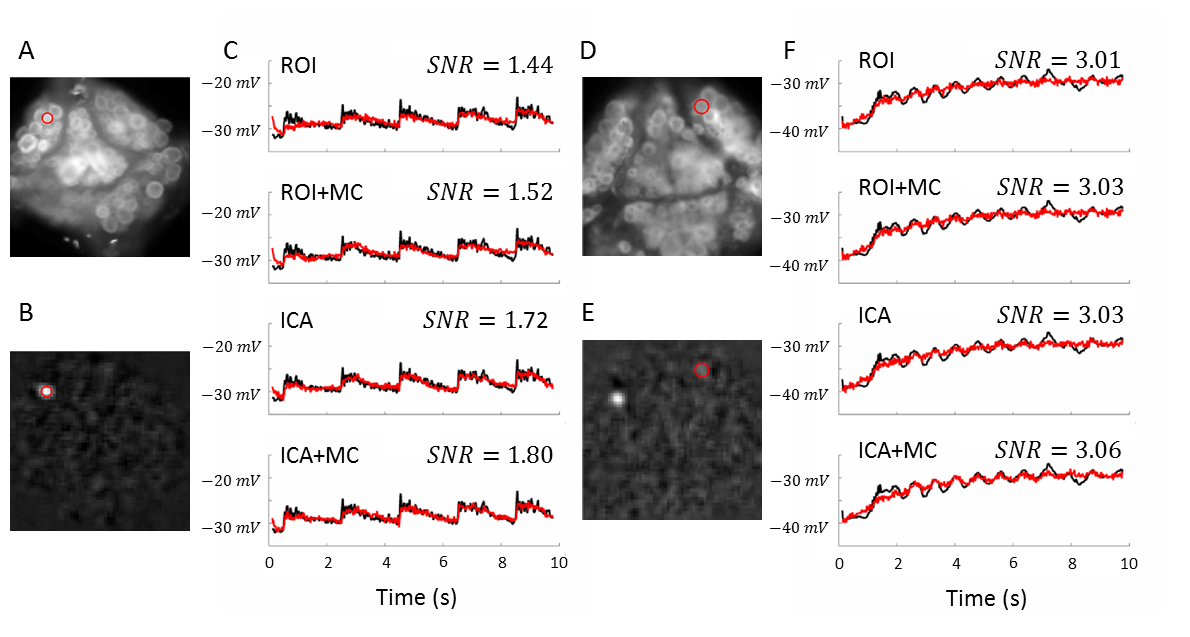}
  \caption[Comparison of ROI and ICA based signal extraction.]{\textbf{Comparison of ROI and ICA based signal extraction.} (A) Raw image with red ROI shown during pre-synaptic stimulation. The optical signal is the average of all pixels within the ROI for each frame. (B) ICA pulls out a component that is manually selected coming from the same cell. The ROI is shown on top to compare localization of component and hand-drawn ROI. (C) The Fit trace is overlayed on the Ideal trace for 4 conditions: ROI only, ROI with Motion Correction (ROI+MC), ICA only, and ICA with Motion Correction (ICA+MC). (D) Raw shown during swim behavior. (E) ICA component. This component shows weight in the bi-lateral pair of cells, as these cells are highly correlated. (F) SNR is compared under the 4 conditions. The 5 mV oscillation are much more clear in the ICA+MC case, even though the SNR increase is fairly small.}

  \label{fig:trace_comparison}
\end{figure}

Figure \ref{fig:trace_comparison} shows some examples of the different extraction techniques under two different conditions (chemical pre-synaptic stimulation, and swimming Fig. \ref{fig:trace_comparison}C, F, respectively). The ROI used for extraction was an oval drawn by hand over the cell being recorded from intracellularly. The ICA component corresponding to the cell was selected using ICM. In some cases, cellular signals are correlated enough to come out as a single component, and occasionally a single IC corresponds to multiple cells (i.e. Fig. \ref{fig:trace_comparison}E). 

\begin{figure}[t]
  \centering
  \includegraphics[width=0.8\textwidth]{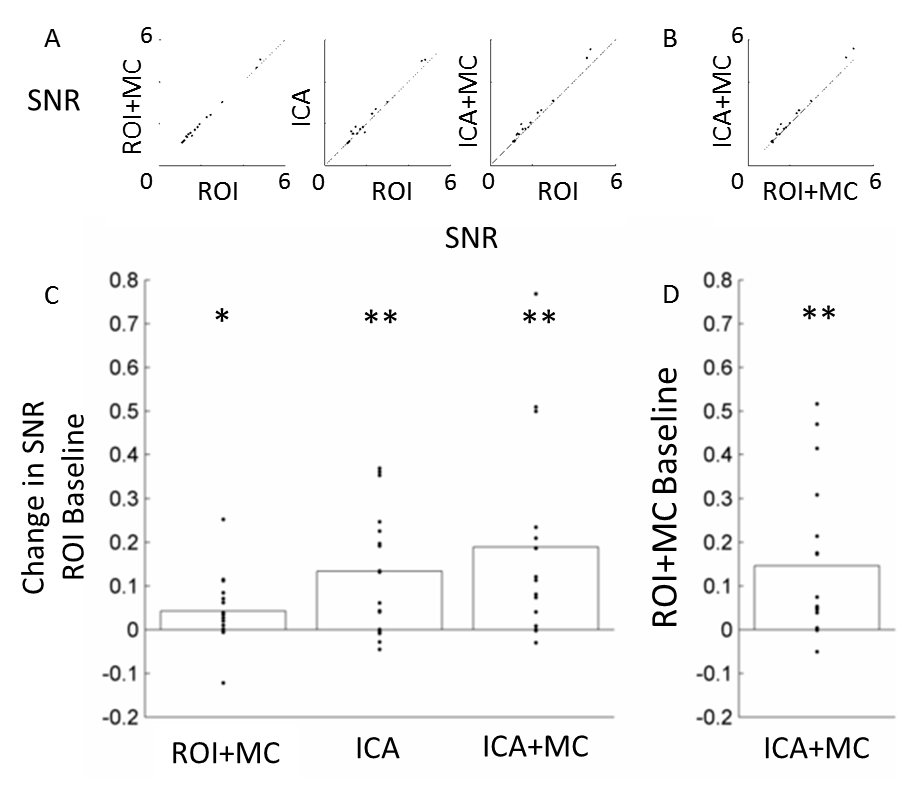}
	  \caption[SNR comparison of different methods.]{\textbf{SNR comparison of different methods.} (A) The SNR for each method is compared against the ROI method. (B) ICA+MC is compared with ROI+MC as the baseline. (C) The difference in SNR for each trace is plotted as black dots, and the average is plotted as the bar graph. Each method significantly increases SNR (ROI+MC: p = 0.0358, ICA: p = 0.0013, ICA+MC: p=0.0024, t-test). (D) Same as C but with ICA+MC compared to ROI+MC (p=0.004).
  }
  \label{fig:snr_comparison}
\end{figure}

For each of the 15 recordings, we compared the SNR across the 4 different methods. Because the SNR of an individual cell varies from cell to cell, we measured the difference in SNR between the different conditions. Compared to the ROI method, ICA improves SNR for almost every recording. Motion correction also improves SNR, and ICA+MC leads to the best SNR results (Fig. \ref{fig:snr_comparison}).

\subsection*{Discussion}
Here we have demonstrated the Imaging Computational Microscope as a tool for automated analysis of large-scale imaging data. This tool can be used to rapidly extract component signals from imaging data and create visualization for real-time experimental feedback. The PCA-ICA extraction has been shown to be a promising technique for analyzing imaging data, and we have created ICM to aid in the rapid use of these algorithms. Further, ICM incorporates user feedback and visualization to overcome some of the short-comings of the PCA-ICA extraction. ICM also includes the ability to perform concatenated trial ICA, which is an important technique to analyze multiple trials of imaging data with consistent components.

We showed that cellular signals extracted with PCA-ICA have improved SNR compared with ROI based methods by measuring the difference between optically recorded and intracellular voltage signals. This improvement is likely because many spatially overlapping components are recorded by each individual pixel, such as signals from neighboring cells or background fluorescence, and ROIs cannot separate these components. ICA, however, can separate spatially overlapping components, which allows artifacts or overlapping cellular signals to be removed from the true source signal.

\subsubsection*{ICA stability}

There are several important theoretical assumptions underlying the use of ICA to extract components from imaging data. The first is that there must be enough data points for the independence of the underlying sources to be accurately computed. This means that there should be many more frames (samples) than there are cells (sources). The second is that ICA assumes that the sources are stationary throughout the data acquisition. To compensate for slight distortions in the spatial location of the image plane, ICM has image registration and motion-correction algorithms that use the visual features of the imaging data to align the sources with the pixels. ICA fundamentally also requires that the signals of interest do not have Gaussian statistics, because ICA cannot assess independence from Gaussian probability distributions. 

The stability of ICA depends on the dimensionality of the data, and in general the higher the dimensionality, the more data needed to extract independent components. Imaging data is extremely high-dimensional, because the dimensionality is proportional to the number of pixels, which is easily larger than 10K for most imaging applications. For some applications, such as EEG (Makeig et al. 1996; Delorme \& Makeig, 2004), ICA can be directly applied to the data because the dimensionality of the data is in the dozens. The primary purpose of PCA is to reduce the dimensionality so that the ICA algorithm can work with datasets with very large dimensionality. This technique can be extended to other modalities of data collection with large dimensionality, such as multi-electrode arrays (J\"ackel et al. 2012) and 3-dimensional acqusition from 2-photon, light-sheet (Ahrens et al. 2013; Freeman et al. 2014), or fMRI.

\subsubsection*{Difference between ICA algorithms}

ICM includes multiple ICA algorithms that can be chosen from. However, these algorithms typically produce similar results, and it is unclear which, if any, algorithm has an advantage. The computation of ICA is an optimization problem, and generally the different algorithms use slightly different methods of estimating the optimization function. The default algorithm, fastica (Hyvarinen \& Oja, 1997), is the fastest ICA extraction and produces consistent results. These results are on the face indistinguishable from the results of other ICA algorithms, but there may be certain advantages in some circumstances. There are many settings that can be tuned to alter the extraction, but usually the most important setting is the choice of PCs to use for ICA. We recommend that the PCs used for ICA be experimented with and the results can be judged by manually viewing the ICs.

Typically, the first $N$ PCs are used for ICA, but occasionally the imaging signal can have noise or artifacts present in the first few principal components. It can sometimes be beneficial to remove the early PCs before performing ICA, especially if they visually appear as pure artifact. However, this demands caution, because the PCs are usually mixtures of components and removing an artifactual PC may also remove some underlying signal.

\subsubsection*{Multi-Trial ICA}
Performing the ICA analysis across multiple trials has some issues because of the random nature of the ICA decomposition. Running the analysis, even on the same data, does not mean that the same components will be extracted in the same order. This is because the initialization of the mixing matrix is random, and the order of the components is somewhat dependent on the initialization of the mixing matrix. We experimented with an alternative solution for multi-trial ICA where we initialized the mixing matrix in one trial using the ICA results from a different trial. This can be performed with ICM using the function \lstinline|init_spatial_guess|. This essentially amounts to providing a guess of spatial locations for expected cellular signals, and the ICA algorithm does find components in similar spatial locations. However, because many cells were tightly packed together, this algorithm often returned neighboring cells as the same component and generally was less consistent than ctICA. Further, this becomes problematic when cells are quiet or show little signal in one or more of the trials. If a cell is not active, then ICA has a much harder time of extracting the signal, because there is less information in the statistics of the pixels. This means that ICA can completely miss a cell from one trial and this will distort the component relationships that are returned in multi-trial analysis.

It is essential for ctICA to have the sources and the pixels aligned across trials, and to help with this we use an image registration algorithm to align the trials. This registration algorithm can only perform an affine transformation on the entire image, and cannot correct for motion of sub-regions of the image or differences in focal depth or z-motion. These types of artifacts will hinder the results of ctICA, but it can still give some good results as long as most pixels over a source remain over the source. More sophisticated image-alignment techniques, especially those with locally deformable registration, can further improve the results of ctICA. Additionally, some of the artifacts from mis-alignment can be compensated through blurring the data in the pre-processing stage. Further, the alignment artifacts are sometimes extracted as a component from ctICA, and these components can simply be ignored.




\subsubsection*{Memory requirements}
The algorithms and data involved in ICM's analysis require very large amounts of memory. Even just a single imaging data set can be gigabytes in size. It is recommended that the software is run with at least 16 GB of memory on the computer, but even this much can be quickly used up by these algorithms. 

\subsection*{Acknowledgements}
This chapter is presented as a manuscript that is in preparation as a publication as Frady, E.P., Kristan Jr., W.B. ``The Imaging Computational Microscope.'' 

\subsection*{References}


\noindent
Ahrens, M.B., Orger, M.B., Robson, D.N., Li, J.M., Keller, P.J. (2013) Whole-brain functional imaging at cellular resolution using light-sheet microscopy. Nature Methods 10: 413-420.
\newline

\noindent
Bell, A.J., Sejnowski, T.J. (1995) An information-maximization approach to blind separation and blind deconvolution. Neural Comput. 7: 1129-59.
\newline

\noindent
Bokil, H., Andrews, P., Kulkarni, J.E., Mehta, S., Mitra, P.P. (2010). Chronux: A platform for analyzing neural signals. Journal of Neuroscience Methods 192: 146-151.
\newline

\noindent
Delorme, A., Makeig, S. (2004) EEGLAB: an open source toolboxfor analysis of single-trial EEG dynamics including independent component analysis. Journal of Neuroscience Methods 134: 9-21.
\newline

\noindent
Evangelidis, G.D., Psarakis, E.Z. (2008) Parametric Image Alignment Using Enhanced Correlation Coefficient Maximization. IEEE Trans. on PAMI [Pattern Analysis and Machine Intelligence] 30(10):1858-1865.
\newline

\noindent
Freeman, J., Vladimirov, N., Kawashima, T., Mu, Y., Sofroniew, N.J., Bennet, D.V., Rosen, J., Yang, C.T., Looger, L.L., Ahrens, M.B. (2014) Mapping brain activity at scale with cluster computing. Nature Methods 11(9): 941-949.
\newline

\noindent
Hill, E.S., Moore-Kochlacs, C., Vasireddi, S.K., Sejnowski, T.J., Frost, W.N. (2010). Validation of Independent Component Analysis for Rapid Spike Sorting of Optical Recording Data. J Neurophysiol 104: 3721-3731.
\newline

\noindent
Hyvarinen, A., Oja, E. (1997) A fast fixed-point algorithm for independent component analysis. Neural Comput. 9, 1483-1492.
\newline

\noindent
J\"ackel, D., Frey, U., Fiscella, M., Franke, F., Hierlemann, A. (2012). Applicability of independent component analysis on high-density microelectrode array recordings. J Neurophysiol 108: 334-338. 
\newline

\noindent
Lee, T.W., Girolami, M., Sejnowski, T.J. (1999) Independent component analysis using an extended infomax algorithm for mixed subgaussian and supergaussian sources. Neural Comput. 11: 417-41.
\newline

\noindent
Lee, T.W., Girolami, M., Bell, A.J., Sejnowski, T.J. (2000) A Unifying Information-theoretic Framework for Independent Component Analysis. Comput. Math. Appl. 31: 1-21
\newline

\noindent
Makeig, S., Bell, A.J., Jung, T.P., Sejnowski, T.J. (1996). Independent component analysis of electroencephalographic data. In Touretzky, D., Mozer, M., Hasselmo, M. editors. Adv. Neural Inf. Process. Syst. 8: 145-51.
\newline

\noindent
Miller, E.W., Lin, J.Y., Frady, E.P., Steinbach, P.A., Kristan Jr., W.B., Tsien, R.Y. (2012) Optically monitoring voltage in neurons by photo-induced electron transfer through molecular wires. PNAS 109(6): 2114-2119.
\newline

\noindent
Mitra, P.P., Bokil, H. (2007) Observed Brain Dynamics, Oxford University Press, USA.
\newline

\noindent
Mukamel, E. A., Nimmerjahn, A., Schnitzer, M. J. (2009). Automated Analysis of Cellular Signals from Large-Scale Calcium Imaging Data. Neuron. 63(6):747-760.
\newline

\end{document}